# Improving Realized LGD approximation: A Novel Framework with XGBoost for handling missing cash-flow data

Zuzanna Kostecka[a,1], Robert Ślepaczuk[b]

[a] University of Warsaw, Faculty of Economic Sciences, Quantitative Finance Research Group, Ul. Długa 44/50, 00-241 Warsaw, Poland, Risk Hub, ING Hubs Poland, 00-351 Warsaw, Poland

[b] University of Warsaw, Faculty of Economic Sciences, Quantitative Finance Research Group, Department of Quantitative Finance and Machine Learning, Ul. Długa 44/50, 00-241 Warsaw, Poland, ORCID: https://orcid.org/0000-0001-5527-2014, Corresponding author: rslepaczuk@wne.uw.edu.pl

**Abstract:** The scope for the accurate calculation of the Loss Given Default (LGD) parameter is comprehensive in terms of financial data. In this research, we aim to explore methods for improving the approximation of realized LGD in conditions of limited access to the cash-flow data. We enhance the performance of the method which relies on the differences between exposure values (delta outstanding approach) by employing the machine learning (ML) techniques. The research utilizes the data from the mortgage portfolio of one of the European countries and assumes the close resemblance for similar economic contexts. It incorporates non-financial variables and macroeconomic data related to the housing market, improving the accuracy of loss severity approximation. The proposed methodology attempts to mitigate the country-specific (related to the local legal) or portfolio-specific factors in aim to show the general advantage of applying ML techniques, rather than case-specific relation. We developed an XGBoost model that does not rely on cash-flow data yet enhances the accuracy of realized LGD estimation compared to results obtained with the delta outstanding approach. A novel aspect of our work is the detailed exploration of the delta outstanding approach and the methodology for addressing conditions of limited access to cash-flow data through machine learning models.

**Note:** This research did not receive any specific grant from funding agencies in the public, commercial, or not-for-profit sectors

---

[1] This paper reflects the author's personal opinions and does not necessarily reflect the views of the Risk Hub, ING Hubs Poland

**Introduction**

One common challenge faced by banks is the presence of data gaps, particularly concerning cash-flow data, which hinder the robust calculation of realized LGD. Although modeling LGD has become a topic that researchers are focusing on recently, there is a substantial gap in the scientific literature for insufficient financial data from the perspective of realized LGD calculation. Analysts often face the challenge of missing cash-flow data, that might be a consequence of scope-jumpers, the customers who changed the portfolio scope. It means that most of their data was not available for some period of their credit history. One says that the data could be accessed eventually, however, some of the processes shouldn't be delayed, such as periodical internal model performance evaluation. Secondly, the precise identification and calculation of the recovery amounts (that are contained in the cash-flows) is considered rather problematic (Ptak-Chmielewska et al. (2023), Khieu et al. (2012)). To address this issue, banks often refer to the delta outstanding method (also called outstanding movement approach), which involves discounting the differences in outstanding balances between reporting periods (ECB, 2024). Delta outstanding approach can be also a solution for researchers, who do not have the access to detailed cash-flow data, due to their high sensitivity and confidentiality. Nevertheless, while this method provides an approximation of realized LGD, it has inherent weaknesses, primarily due to its omission of detailed cash flows.

This problem can be addressed by finding more complex dependencies and patterns between available data and customers' LGD. We propose to utilize the eXtreme Gradient Boosting (XGBoost) model to improve the performance of delta outstanding approach approximation for the realized LGD. Our aim is to show which variables can be included in the model and, through the analysis of performance metrics such as Mean Absolute Error (MAE) and Mean Squared Error (MSE), show to which extent one can benefit from usage of ML techniques.

The main contribution of this research to the existing literature and the novel aspects of approaching the results of delta outstanding method are: (1) quantifying and incorporating of non-cash-flow related variables, and (2) introducing the ML model to uncover hidden patterns among variables, which is an opposite to delta outstanding approach that concerns only the discounted arbitral change of outstanding (3) addressing the conditions of limited access to the data (especially cash-flow related data).

Our dataset, obtained from one of financial institutions, contains the information about retail mortgage portfolio of one of the European countries. It spans from 2008 to 2019, with monthly frequency, providing a rich and comprehensive basis for our analysis. Apart from borrower-specific data and credit information, the dataset is also enriched with macroeconomic variables, which are reported in quarterly frequency. The data include only defaulted customers, which according to the literature may lead to imbalanced sample (K. Li et al. (2021), Fan et al. (2023), Gürtler and Zöllner (2022)). Further discussion on not-perfectly balanced data is going to be described wider in the *Literature review* section in *Prediction models* subsection.

In the context of write-offs, it's important to note that the delta outstanding method fails to recognize the underlying cause of changes in outstanding balances due to accounting write-offs. Consequently, there's no evidence available to determine whether the decrease in amounts is attributable to repayments or to the write-off process (the accounting process of writing off some amount because it has been given up by the bank in due to the poor solvency of the client). Therefore, the important feature of the mortgage data, as well as this specific portfolio, are rare cases with accounting write-offs, but also additional drawings, undrawn amounts and prepayments. These factors contribute to the suitability of our sample for delta outstanding approach, ensuring the reliability of obtained results.

Accurately approximated realized LGD for the model monitoring purposes is the substantial factor, that contributes to the model (predicted LGD) performance and its conservativeness assessment (Prorokowski, 2022). Inaccurate predicted LGD can lead to misallocation of capital reserves, potentially resulting in insufficient provisions for credit losses (Spuchľáková and Cúg (2015b), Hurlin et al. (2018)). This can undermine a bank's ability to absorb unexpected losses and meet regulatory capital requirements, posing systemic risks to the financial system. Given the significance of LGD calculations, our study aims to contribute to the advancement of credit risk analytics by proposing a refined approach. By leveraging ML techniques, we seek to develop a model that enhances the accuracy of realized LGD estimation, particularly in cases where traditional methods fall short. In this research we attempt to answer the following research questions:

(RQ.1) *If approximation of realized LGD, using method relied on differences in outstanding balances between reporting periods, can perfectly reflect the reality (assuming the true realized LGD is the result of cash-flow approach)?*

(RQ.2) *Does inclusion of non-cash-flow related variables, such as macroeconomic data, customer's final status, or predictors built on fundamental variables (default date, reporting date, outstanding balance), lead to improved accuracy in approximating the realized LGD?*

(RQ.3) *Is there any subsampling method that improves the performance measures (MSE and MAE) for XGBoost model?*

The structure of the rest of paper follows the below sequence. (1) In section *Literature Review* we introduce the substantial assumptions and definitions, also going through academic and business publications on LGD modeling topic. (2) Section *Methodology* represents insights of calculation of the LGD based on three different approaches: cash-flow approach, delta outstanding approach and delta outstanding enhanced with ML model. (3) Section *Results* includes the performance comparison of different approaches. (4) In the last section, *Conclusions*, we highlight our findings and summarize our paper.

## 1. Literature review

In this section, we propose a broad overview of academic papers on default status and mortgage loans, while both are the object used for this analysis. Then, we summarize existing literature on topic of the importance of Loss Given Default (LGD) calculation, and conclude on potential challenges regarding this parameter. Finally, we present different approaches, both on model and variables selection, which enhance the calculations performance.

### 1.1 Default definition

The precise definition of the default event, here and after referred to as the Definition of Default (DoD), reveals a lack of uniformity and comprehensive standardization across the different institutional frameworks (Spuchľáková and Cúg, 2015). According to Moody's (2011), "Default is defined as failure to make scheduled principal or interest payments". Followingly, in 2016 the European Banking Authority (EBA) published the guidelines on defining a default (with a reinforcement in 2021), called new definition of default (NDD) (Prorokowski, 2022). From the perspective of this research, the portfolio's realized LGD may vary whenever the default definition changes, as the default period per customer can potentially be different. Thus, as Barisitz (2019) notes, it is important to hold the crucial aspect (also called "primary elements") standardized across different jurisdictions:

- count of past due days: non-performing exposure is identified after 90 days past due,
- materiality: only material exposures are considered for above condition,

- unlikeliness to pay: the debtor, who is assessed as unlikely to pay without the full realization of collateral is classified as non-performing.

The above conditions provided in the Capital Requirements Regulation (CRR) and interpreted by the European Banking Authority (EBA) are consistent with the Basel Committee on Banking Supervision's (BCBS) definition of default and with the Institute for International Finance's credit quality classification. Nevertheless, Nehrebecka (2018) raises, that primary elements are still leaving a considerable margin for unconfined interpretation.

While the determination of default status is one matter, one can also focus on the default reasons, which apart from their delays in payments, might also refer to the legal or solvency matters (Kao, 2000). Even though the default reason is rather understood as a risk driver than the loss predictor, we should mind that based on this information we might reveal insightful patterns related to the chances, that the customer will be able to cover their outstanding amount with the collateral. For example, according to Witzany (2017), the bankrupt clients are rather going to expose the bank on the loss of most of their outstanding amount. Khieu et al. (2012) provides with example of default reason significance by highlighting the fact that bankruptcy of the borrower is a common reason to revise the terms of the loan contract.

## 1.2  Mortgage Loans

Although the majority of researchers are focusing on precise estimation of LGD rather than handling the data availability for realized LGD, their conclusions supply us with multiple insightful, field-specific characteristics for mortgage loans. The practice of selecting mortgage loans for analysis is widely employed by researchers due to its profound impact on financial stability, as highlighted by Park and Bang (2014). An important part of the mortgages loans are commonly applied collateralizations, that contributes to the stability of financial systems by providing lenders with a form of security against borrower defaults. According to Tong et al. (2013), after a mortgage loan defaults, one potential outcome is that the property undergoes repossession by the bank. This practice allows banks to mitigate the risks associated with lending, making them more willing to extend credit to borrowers.

Another advantage of choosing the mortgage data for the LGD modeling has been outlined by Ptak-Chmielewska et al. (2023). The long workout period observed in mortgage loans supplies our dataset with a comprehensive default history and helps to uncover not only point in time patterns but also historical trends that might be beneficial for the ML model.

Leow and Mues (2012) or Hurlin et al. (2018) papers provide evidence supporting the interchangeability of "outstanding" and "exposure" terms, particularly in the context of mortgage loans. This equivalence stems from the common occurrence in mortgage loan scenarios where minor additional drawings lead to both parameters being practically equivalent.

## 1.3 Loss Given Default

When assessing credit risk, critical metrics such as Exposure at Default (EAD), Probability of Default (PD), and Loss Given Default (LGD) are derived through both empirical observation and predictive modeling. These parameters serve as inputs in the calculation of regulatory capital requirements (European Central Bank, 2021). It refers to the Advanced Internal Rating Based modeling standards (AIRB), that allows banks to use their own resources to model the risk parameters (Ptak-Chmielewska et al., 2023). Predicted values are utilized in computing capital reserves, ensuring banks hold reserves corresponding with their risk exposures (Hurlin et al., 2018). Regulatory Capital prescribed in the Basel Framework, ensures banks maintain sufficient capitalization to absorb potential or unexpected credit losses, in other words counterbalance potential liquidity issues resulting from defaulted customers within the portfolio (Chiodo and Hasan (2013), European Central Bank (2021)). Conversely, realized values of different risk parameters, (including realized LGD) derived from historical data play an important role in calculating the Risk Weighted Assets (RWA) (Liu (2017), Akkizidis and Kalyvas (2018)). The accurately calculated realized values of risk parameters are also substantial for validating and recalibrating predictive models, and thereby enhancing their accuracy and reliability (Park and Bang, 2014).

According to Hurlin et al. (2018), the LGD estimates enter the capital requirement formula in a linear way and, as a consequence, the estimation errors have a strong impact on required capital. The determination of the LGD is primarily contingent upon the ratio of the actual loss to the total exposure as of the borrower's default reporting date. Tong et al., (2013) defined the LGD formula as proportion of the outstanding loan that will be lost in the event of a default. Since the value is non-zero only given default, one can write down the formula for computed LGD in the **Equation 1** as follows:

$$LGD_{comp} = \frac{EL}{EAD} * default\_ind \qquad (1)$$

where: EL is an Economic Loss, EAD stands for Exposure at Default, default_ind is the binary default indicator, which is equal to 0 when the borrower is not in the default and equal to 1 when the borrower is in default (EBA (2018), Hurlin et al. (2018), Xuan et al. (2018)). In other words, the aim of the LGD modeling is to accurately estimate the lender's loss in relation to the total outstanding in conditions of the defaulted loan (Leow and Mues, 2012, Spuchľáková and Cúg (2015a), Sproates (2017)). In cash-flow approach we calculate EL, i.e. discount the cash-flows which include both recoveries and costs. Recoveries, that lower the bank's loss consists both of cash - actual cash flows collected from defaulters during the workout period and non-cash – repossessions of collaterals.

## 1.4 Delta outstanding approach

The most dependable, frequent and preferred by the regulators method for calculating the economic loss for realized LGD is rooted in discounted value of realized cash flow analysis, called workout approach (Ptak-Chmielewska et al. (2023), Hurlin et al. (2018), EBA (2018)). Even though the discounting cashflows reflects the reality precisely, banking industry has developed different methods of approximation the LGD: discounting change in balances (delta outstanding approach), or discounting write-offs (Australian Prudential Regulation Authority (2024), ECB (2024)). Importantly for this research, we outline that European Central Bank (ECB) permits the utilization of an alternative approximation based on the outstanding deltas in certain applications, such as model monitoring.

The delta outstanding approach is being frequently utilized in banking industry in conditions of limited access to the borrower's cash-flows, or according to ECB (2024) when recoveries are not directly observed. This can be a case especially when the borrower is a scope-jumper, which in other words stands for the borrower who for certain period has been out of the analyzed portfolio scope. This behavior might cause significant gaps in the data and might result in insufficient information for calculation of realized LGD through the cash-flows method, and consequently, impossible assessment of the estimated LGD model.

Nevertheless, the weakness of the delta outstanding approach is that it arbitrarily relies on the changes in outstanding amount. Gürtler and Hibbeln (2013) highlights that neglecting workout costs leads to misestimation of LGD. Moreover, according to Hurlin et al. (2018), workout costs are rarely included in empirical researches due to problematic nature of capturing them.

On the other hand, EBA (2024) emphasizes that institutions should put their best effort to adequately replicate the cash-flows and document the substitutional approach in a way that leads to a clear and consistent justification of the treatment. This supports the potential value of our novel approach that enhance LGD estimation without relying on cash-flow data, offering inclusion of the hidden patterns in data as a predictive factor, inclusion of highly available variables and promising avenues for advancing research in this domain.

**1.5   Selecting variables in Estimated Loss Given Default models for mortgage loans**

Although the approach and formula for the realized LGD can be similar among different portfolios and industries, the estimated LGD calculation should take into account the field-specific factors. Given that we estimate the realized LGD using a supervised model and thus utilize predictive methods, our variable selection can be guided by the existing models used to estimate LGD.

Leow et al. (2014) suggest that LGD is correlated to the economy and they conclude, that macroeconomic variables are able to improve the model both for private and corporate mortgage loans. According to multiple publications (Leow et al. (2014), Park and Bang (2014), Xuan et al. (2018), Hurlin et al. (2018)) it will be relevant to search for the loss severity patterns through the macroeconomic indicators such as the House Price Index (HPI), change in unemployment rate, economic growth measured with the Gross Domestic Product (GDP). Tong et al. (2013) in their analysis considers discounting factors, because it aims to assess precisely more complex patterns and correlations hidden under the customer's history. EBA (2018) also identify discounting factors with the interest rates, which for European countries can be derived from Euro Interbank Offered Rate (EURIBOR) and a 5% add-on. They propose this approach explaining it as simple and contributing to increased comparability of LGD estimates. Importantly, the discounting factor used for LGD calculation is focused on the uncertainty in the recovery process and the time value of money.

Apart from highly accessible macroeconomic variables, our model should dive into more loan specific variables. Ross and Shibut (2015) concludes, that workout period is an important indicator of the loan servicing (for effective servicing of the loan, the assets cover properly the loss and prompt to the cure status) and origination quality (weaker credits tend to prolong the workout period and are indicative of lower origination quality). Referring to Kaposty et al. (2020) results of the research, duration of the workout period may have a significant impact on the LGD estimate. Based on the Tong et al. (2013) research, we can also conclude that customers, who already defaulted in their past should be identified in the model. According to Hurlin et al. (2018) research, also the EAD, even if not directly correlated with LGD, may bring a predicting power to the model. However, they also warn about the large differences in EAD between borrowers, that may impact final regulatory capital forecast error.

Behavioral variables, that are more commonly used for the PD modelling, are also included in scientific research as a meaningful indicators of loss severity (Ross and Shibut (2015), Li et al. (2023)). In most cases, selected behavioral variables (e.g. borrower occupancy, property types), are not used to be incorporated alone, without both the macroeconomic indicators and loan characteristics. Although, our dataset limitations include lack of behavioral variables.

In summary, in the domain of mortgage LGDs, current practice and literature revealed three distinct categories of relevant predictor variables: loan attributes, behavioral elements, and macroeconomic indicators (Xuan et al., 2018). Nevertheless, the limitation of our dataset includes omission of behavioral data. We will attempt to compensate absence of behavioral data with access to whole default history, that gives us a wide field for extraction of the hidden patterns and trends in customer tendency to repay the loan.

## 1.6 Prediction models

Hurlin et al. (2018) highlights the high heterogeneity among banks and academics using the AIRB approach. They observe that the researchers utilize the parametric regression models (linear regression, survival analysis, fractional response regression, inflated beta regression, or Tobit models) or non-parametric (regression tree, random forest, gradient boosting, artificial neural network, support vector regression, etc.) for LGD. They also noticed that parametric models dominate over the non-parametric models in terms of interpretability. On the other hand, non-parametric models are in favor of predictive power, which is the main goal for our research. Simultaneously, the non-parametric models are also commonly recommended in other academic papers (Qi and Zhao (2011), Bastos (2010), Loterman et al. (2012)). Finally, Leow et al. (2014) summarize that among different academic models, the OLS has the biggest prevalence.

Tong et al. (2013) noticed that the simple regression approach will bias the prediction whenever the sample is strongly imbalanced towards LGD equal to 0. They propose a zero adjusted gamma model, which incorporates probability of a zero loss and the loss amount given that a loss occurs simultaneously. Xuan et al. (2018) noticed, that there is not only a dilemma, if the borrower will cure or expose on further losses, but also if the collateral value will cover the outstanding loan amount. They propose a three-step selection approach with a joint probability framework for default, cure and non-zero loss severity information. Leow and Mues (2012) focus on distinction for selecting the model, depending on its nature, stressing that inclusion of two stage model is beneficial for the mortgage loan, while OLS regression is appropriate selection for the personal loans. They are providing with the evidence that accuracy of LGD estimation can benefit from the Haircut Model, where the haircut represents the discount factor to be applied to the estimated sale price of the property. Indeed, the model incorporates the probability of the repossession, and the sale price of a repossessed property may undergo. Those two combined results in the estimation of the LGD. Nevertheless, this approach might enlarge the issues related to the data availability, since even 93 variables have been used for the modeling part. Ptak-Chmielewska et al. (2023) address another problem, which is precise identification and calculation of the recovery amounts, especially for the not-resolved cases. As a solution, they apply the kNN model, trained on resolved cases with a possibility to be applied on not-resolved cases. Aforementioned method offers a promising alternative to our approach, but it relies on incomplete recovery data, unlike our case, which assumes a lack of access to cash-flows (both recoveries and further costs).

Finally, the imbalance of the target variable in LGD modeling is rather associated with the classification problems (Baesens and Smedts (2023), Coşer et al. (2019), Merćep et al. (2020)). Moreover, as noted by Xia et al. (2021), LGD modeling is commonly converted to the regression problem, as is the case in this study. Therefore, our focus is rather on identifying the most effective model for regression problems similar to one represented by our case, than seeking solutions specifically for imbalanced target variables. G. Li et al. (2022) noticed that in regression problem XGBoost outperformed other models (Random Forest, Ada-Boost and GBRT) when applied on mixed-type input data and skewed distribution of the output variable. Fan et al. (2023) emphasize several advantages of XGBoost models in context of LGD: preventing overfitting, comparable results to other models within a shortest time with fewer computing resources, or finally, in case of decomposing the LGD prediction into two classification problems and one regression problem, better predictive ability.

## 1.7 Summary of literature

Definition of default is not problematic from our perspective, because we work only on defaulted cases, and its role is rather for the context. Also, the definition of realized LGD is consistent among different sources with respect to the cash-flow approach. The challenging aspect is the definition of the delta outstanding procedure, which we find mostly general and limited in terms of scientific literature. If it comes to the modeling of LGD, we observed a tendency of application the two-stage models, while most frequently first stage referred to the question if the economic loss occurred or not as a classification problem. Secondly, the LGD value estimation could be both put into classification problem or converted to regression problem. LGD estimation as a regression problem wasn't the most popular approach in the literature, but while it was present, the conclusions led to recommendation of usage of XGBoost model.

## 2. Methodology and data

In this section we focus on providing substantial information about the dataset. Then, we present the procedures on calculating of LGD with cash-flow approach and delta outstanding approach. Lastly, we describe the novel approach of managing the low accessibility to detailed cash-flow data in calculating realized LGD, utilizing the ML techniques.

### 2.1 Dataset description

The data are reported in monthly frequency and spans from the year 2008, including more than 10 years of observation for total of 1891 borrowers. Since only in-default observations are included in the dataset for analysis, their LGD (excluding last observation per single default) is non-zero. In several cases, particularly for borrowers with final status 'cure' (explained further in subsection *Delta outstanding approach combined with the XGBoost model*), the LGD in-default is small or close to zero. The distribution of realized LGD values is represented in Figure 1.

**Figure 1.** Distribution of realized LGD with U-shaped pattern.

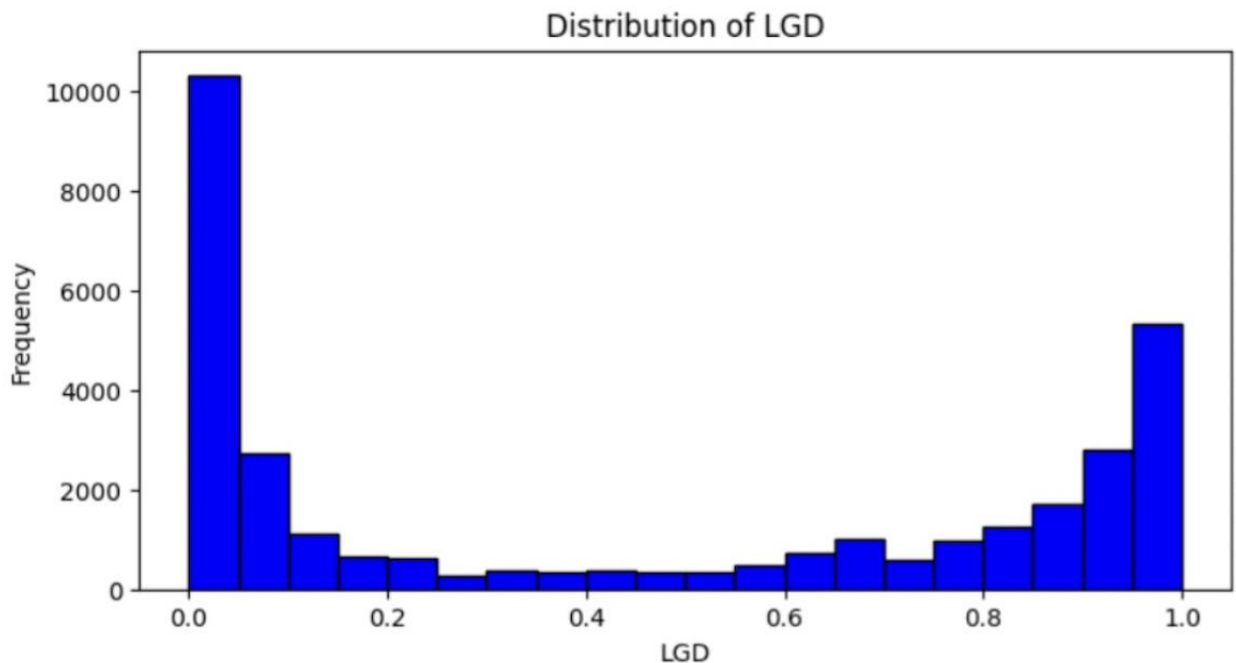

Note: The plot is created based on the sample with capped observations LGD < 0 (see, Equation 3) and excluded observations with LGD > 1. Based on own work.

The dataset employed in this study is aggregated at the level of customer ID, default date, and reporting date. We recognized in our data 104 borrowers who experienced multiple defaults (with the time period between the distinct default above 3 months). In cases where customers exited default within a probation period of less than 3 months, the consecutive default periods are consolidated into a single entry.

The duration of default varies among customers, ranging from less than a month (considered as 0 since each period equals one month) to a maximum of 130 periods. This results in 56k observations in total. Figure 2 illustrates the distribution of default duration, with multiple-default occurrences treated as distinct instances:

**Figure 2.** Distribution of distinct defaults duration.

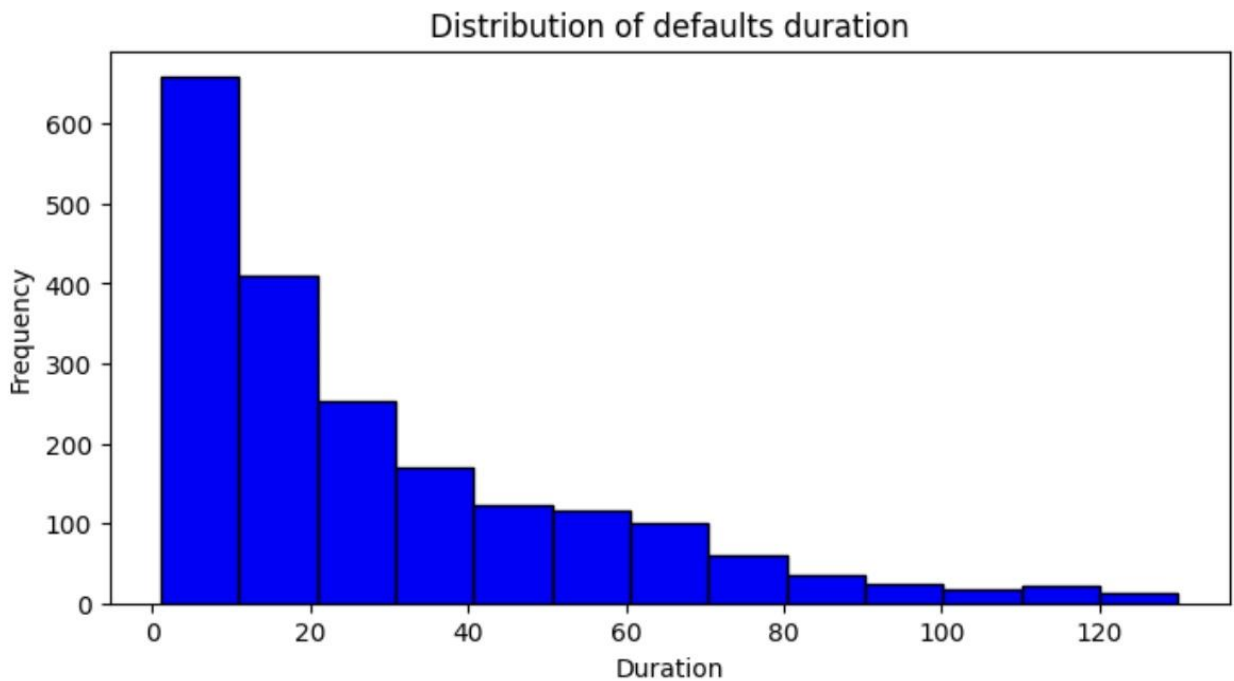

Note: For borrowers who defaulted multiple times (with period between defaults >3 months), their defaults are treated as separate inputs for the plot. Duration is denoted in months. Based on own work.

As already mentioned in the introduction, accounting write-offs do not contribute significantly to our data (53 borrowers with non-zero write-offs during the whole default history), similarly as additional drawings. While the assumption regarding additional drawings remains a theoretical statement referred from literature of mortgage data, it's important to note that our dataset is limited in this context, as it does not contain such a variable. Nevertheless, we have information about the write-offs, which distribution illustrated in Figure 3.

**Figure 3.** Distribution of write-off amounts in the sample.

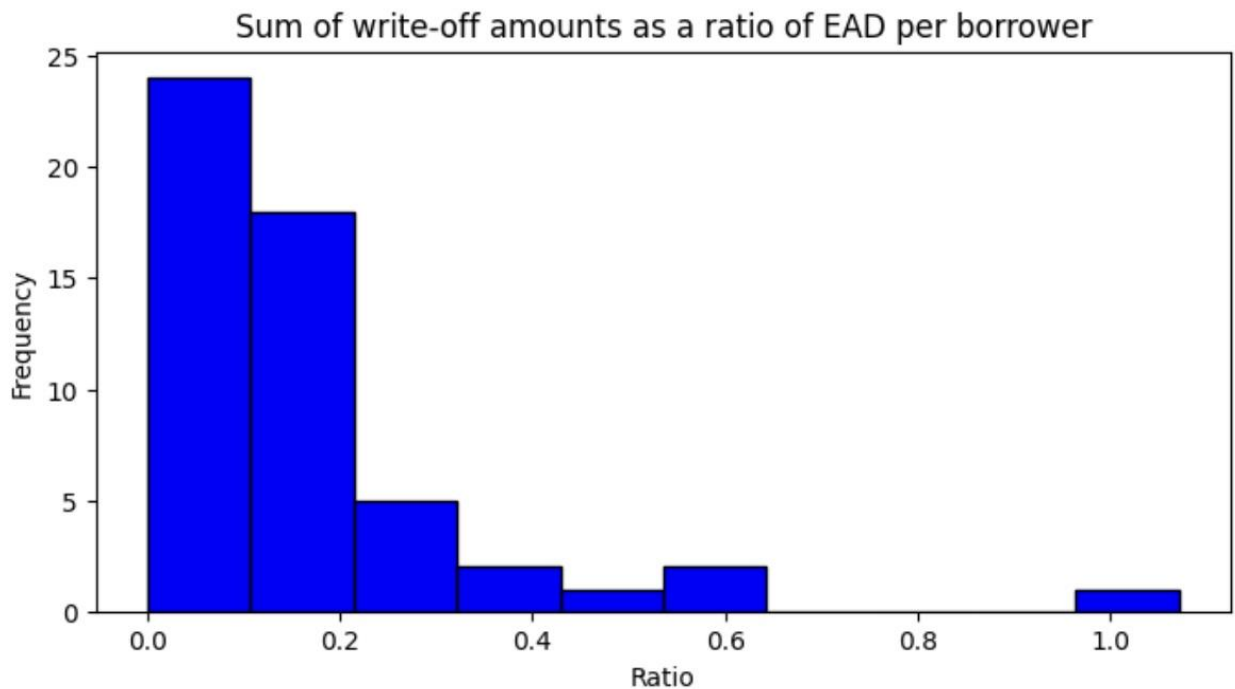

Note: Ratio means the sum write-off amount of a single default during its whole period divided by EAD amount. Ratio equal to 1 means that the whole outstanding at default amount has been written off (either once or gradually) and it concerns only one borrower in our dataset. Based on own work.

We observe that the majority of write-offs are below 20% of the total borrower's EAD. In one case the write-off was equal to the total loan value. The cases where the write off has been non-zero are present among less than 3% borrowers.

## 2.2 Cash-flow approach

The cash-flow approach follows the Equation 1 introduced in the literature review. Referring to that, we can precise the definition of economic loss (EL) using below Equation 2:

$$EL = EAD - \sum_{t=1}^{T} \frac{recovery_t}{(1+r_t)^t} + \sum_{t=1}^{T} \frac{cost_t}{(1+r)^t} \qquad (2)$$

As observed by Hurlin et al. (2018), the theoretical LGD ranges between 0% and 100%. Nevertheless, Miller and Töws (2018) shows, that incorporating the impact of discounting factors during the workout period, may result in LGD higher than 100%. We refrain from implementing any adjustments to these factors, as their impact is minimal and they are relatively easy to interpret (due to factors such as legal fees, recovery delays, and economic fluctuations affecting asset valuations). Moreover, Leow et al. (2014) observed that after discounting the cash-flows, some values might be also negative. In our dataset, we faced a small number of observations with negative LGD. We implemented the approach recommended by EBA (2018), which is capping the negative LGD to 0 (less than 2% of the observations required the application). The approach is supported by better interpretability and lower bias for the model. Equation 3 below represents the mathematic interpretation of the transformation we have applied:

$$LGD_{model} = \max(0, LGD_{comp}) \qquad (3)$$

## 2.3 Delta outstanding approach

The delta outstanding approach is an alternative for calculation of the LGD honored by ECB. The procedure is referred to as differences in outstanding balances between reporting periods (EBA, 2024). Methodology described by EBA (2024) highlights the importance of adequate replication of recovery cash-flows, including the correctness of used interest rates and fees even from the period before the default. Since we omit the impact of additional drawings, which is also part of the broader applicability of the research, there is no further methodology description for the approach in the *ECB guide to internal models.* We can use recommendations proposed for the calculation of LGD (definition of outstanding, definition of discounting factor, treatment of cured cases) to align with the most dependable methodology.

In this research, delta outstanding approach utilizes the EURIBOR interest rate for discounting. The discounting factor is defined as in Equation 4 below:

$$disc_{factor} = \left(1 + \frac{interest\ rate}{12}\right)^t \qquad (4)$$

where *t* is the number of months between the reference date (to which we discount) and reporting date (from which we source the outstanding amount for the discounting). Moreover, we follow the naming convention, where the reporting date is the date of the observation, and it is to reference date or later. The data per customer is limited from their default date until the out-default date. Each observation is factorized as follows:

1. Define outstanding balance, which referring to EBA (2024) is the accounting value gross of the credit risk adjustment. Initialize the variable os_prev to represent the outstanding balance on the previous reporting date. If the reporting date is equal to the default date, set os_prev to 0.
2. Expand the dataset to include in-default reference dates. Each reporting date is considered as a reference date, and for each reference date, all available reporting dates are separately analyzed. This expansion results in multiple rows for each reference date, ensuring thorough examination of the data, that means if we have n periods (months, reporting dates) for a customer, they will be extended to n! rows. The procedure is presented in Algorithm 1 below.

---

**Algorithm 1.** Expanding the data for delta outstanding approach.

1: CREATE TABLE expanded_table AS
2:     SELECT a.*, b.reference_date AS reporting_date
3:        FROM initial_table AS a
4:     LEFT JOIN initial_table AS b
5:        ON a.id = b.id
6:        AND a.reference_date <= b. reporting_date;
7:    ORDER BY a.id, a.reference_date, b.reporting_date

---

3. Calculate delta outstanding values using *os_prev*, where *delta_os* for *reference_date=reporting_date* is always equal to 0. Discount the *delta_os* values from reporting date to reference date using the discounting factor.

4. Calculate cumulative sums of discounted delta outstanding values. Each cumulative sum initializes per new *reference_date*.

5. Calculate the Expected Loss (EL) by subtracting the cumulative sum from outstanding amount of corresponding reference date (*os_ref*).

6. Determine the Realized Loss Given Default (RLGD) by dividing EL by *os_ref*.

7. Filter the dataset to retain only the rows where, for each reference date, the corresponding reporting date is the maximum value. The last record per reference date contains the net recoveries from the reference date to the out-default date.

## 2.4  eXtreme Gradient Boosting models

The XGBoost algorithm, proposed by Chen and Guestrin (2016), is a member of the gradient boosting family models. Apart from XGBoost, the most popular implementations within this family of models are: Gradient Boosting Machines (original form of gradient boosting), LightGBM or CatBoost**.** Gradient boosting is a powerful ensemble learning technique that combines the predictions of multiple individual models (typically decision trees) to create a stronger joint model (Binder et al., 2014). In more detailed words, the idea is to sequentially train a series of weak learners, each one focusing on the mistakes made by its predecessors. When the weak learner is characterized by a poor performance, the algorithm assigns higher weights to the misclassified or poorly predicted instances, thereby prioritizing them in subsequent iterations. By focusing on the most challenging cases, boosting aims to gradually improve the overall predictive accuracy of the model. The final prediction is obtained by combining the predictions of all weak learners, with each one contributing more weight to the final prediction based on its performance (Dong et al., 2022). According to Guo et al. (2020), the formula for XGBoost can be represented as:

$$\hat{y}_i = \sum_{k=1}^{K} f_k(x_i), f_k \in F \qquad (5)$$

where $\hat{y}_i$ stands for the predicted value, $K$ is the number of trees, $f_k(x_i)$ is the function of input in the $k$-th tree, and $F$ is the set of all possible regression trees. They state that as the XGBoost algorithm uses the gradient boosting strategy, it adds one new tree at a time and continuously improves the previous results by fitting the residuals of the last prediction. This process can be represented by modifying the function of input from Equation (5) as the new function $f_k(X, \theta_k)$, where $\theta$ stands for the residuals. Once K trees have been trained, each feature of the prediction samples will correspond to a leaf node in each tree, with each leaf node associated with a specific score. The final prediction value for the sample is obtained by summing the corresponding scores from all the trees. The flowchart of the XGBoost process is showed in Figure 4.

**Figure 4.** XGBoost algorithm structure

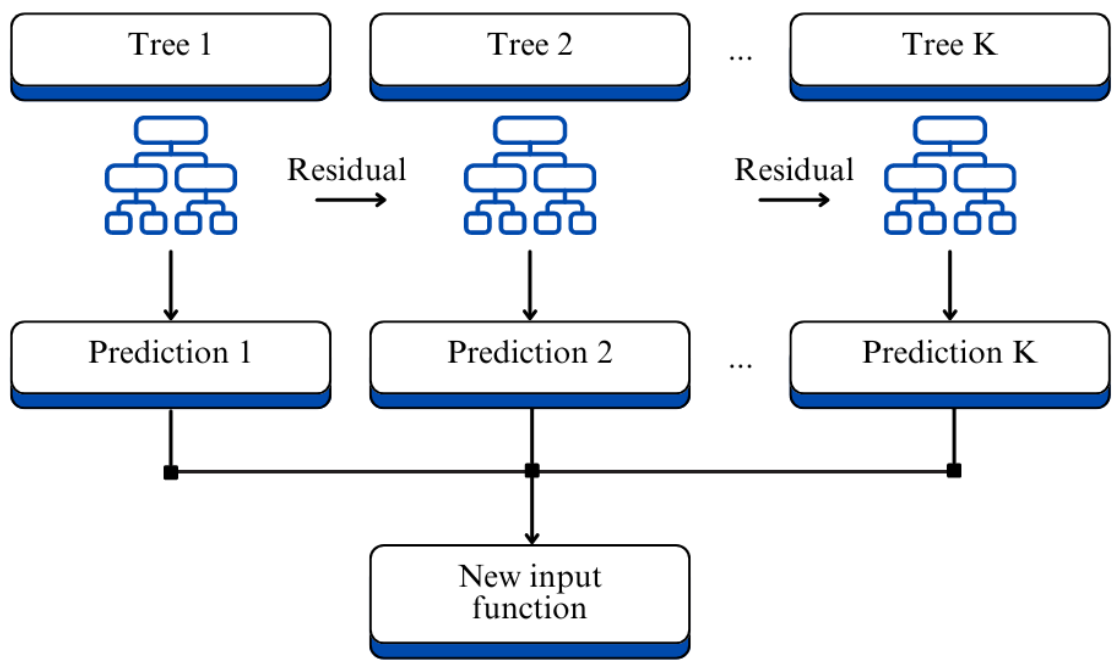

Note: Presented XGBoost algorithm flowchart refers to the process performed on the training sample.

XGBoost has been proposed as a highly optimized, scalable machine learning method. It means that it was designed to increase the ability to effectively handle increasing amounts of data without a significant increase in computational resources or time. The algorithm differs from traditional gradient boosting methods in its optimization techniques, including a more efficient handling of sparse data and a regularization term in the objective function, which helps prevent overfitting (Chen and Guestrin, 2016). Followingly, the loss function consists of two parts, where one is the fitting error (e.g. MAE, MSE) and second is the regularization term, that penalizes the model complexity.

According to authors of the XGBoost, algorithm performs outstandingly for the real-world data that includes frequent zero entries or artifacts of feature engineering such as one-hot encoding, that are also part of dataset of our research.

## 2.5 Variables selection for XGBoost model

In our modeling approach we refer to the benchmarking methods that, as Hurlin et al. (2018) noticed, are utilized by banks and academics. Main objective of the procedure is to compare several approaches and select the most accurate based on the selected and interpretable performance metrics, such as Mean Squared Error (MSE) and Mean Absolute

Error (MAE), also used in research by i.a. Kaposty et al. (2020). Benchmarking in this study involves evaluating the performance of three approaches: the delta outstanding approach, along with two variations of XGBoost models. The first XGBoost model is applied to the entire training sample, while the second approach involves training two separate XGBoost regressors on subsets of the sample created based on the final status of the borrower.

It is important to understand the different status of customers, since based on them the data might benefit from dividing into two different samples with separate regressors (Xuan et al. (2018), Tong et al. (2013), Ptak-Chmielewska et al. (2023)). Apart from estimating the XGBoost model on the whole sample, independently on the final status of the customer, we will estimate the XGBoost on grouped data. First sample of grouped data approach, 'noloss', includes (1) customers who are cured and (2) those who exit without loss. (1) Cured customer was defaulted but he went out of default (is now performing again) (Wood and Powell, 2017). (2) Exit without loss on the other hand means that one has been written-off, and the bank recovered 100% of the outstanding amount. In this context, write-off refers to the accounting treatment of the loan, indicating that the bank does not expect to recover any further payments from the borrower. It's essential to note that the term "exit without loss" implies that the bank has managed to recover the outstanding amount through other possible channels, such as selling the collateral, negotiating a settlement with the borrower, or using other recovery mechanisms. In the second sample, 'loss', we concern two more final status of the borrower: (3) "not resolved" and (4) "exit with loss". (3) Not resolved status does not mean that the bank incur losses yet, but since the best practice in safe credit risk management is heading towards conservative approach and for that reason, we classified this group to a riskier group. It can also be called the dragging case. (4) Exit with loss, as the name suggests, is the situation where the debt has been written-off and the bank did not recover 100% of the outstanding amount.

The model utilizes 19 candidate predictors listed in Table 1. Several variables contain the information about the customer out of the observation point in time (e.g. during the borrower's default the final status is already known). It is because our model is built for the realized LGD approximation, that is approached similarly as the backtests. Such cases concern that the analyst already have the general data about the customer, reaching from the historical period, until the end of the observation period. This approach brings additional information to the model and referring to Tong et al. (2013) Basel II Accord interpretation, is also recommended.

**Table 1.** List of financial and macroeconomic variables for modeling the loss amount with LGD

| Variable/ Level | Noloss | Loss | Total sample |
|---|---|---|---|
| Unsecured_recovery_interest | -0.078 | -0.061 | -0.159 |
| Secured_recovery_interest | . | 0.005 | 0.004 |
| Cover_value_index | 0.119 | -0.075 | -0.027 |
| EAO | 0.205 | -0.026 | 0.123 |
| Discount_rate | 0.239 | -0.039 | 0.073 |
| Os_delta | -0.051 | -0.033 | -0.056 |
| RLGD_os | -0.159 | 0.889 | 0.546 |
| Default_duration | 0.595 | 0.007 | 0.511 |
| Default_start_reason_90DAYS | 0.050 | -0.033 | 0.042 |
| Default_start_reason_BANKRUPT | . | 0.006 | 0.009 |
| Default_start_reason_FORBPERIOD | -0.039 | 0.003 | -0.112 |
| Default_start_reason_FRAUD | . | 0.115 | 0.161 |
| Default_start_reason_RESTR | 0.031 | 0.010 | -0.033 |
| Default_start_reason_UNLIKEPAY | -0.043 | -0.072 | -0.064 |
| GDP | -0.125 | 0.087 | 0.067 |
| Employment | -0.188 | 0.204 | 0.202 |
| HPI | 0.181 | -0.200 | -0.175 |
| Repayment | -0.003 | 0.011 | 0.001 |
| Redefault | 0.112 | -0.107 | -0.235 |

Note: Unsecured_recovery interest – interest rate for discounting of unsecured recovery; Secured_recovery interest – interest rate for discounting of secured recovery; Cover_value_index – security value; EAO – Exposure at reporting date; Discount_rate – discount rate calculated as EURIBOR + 5%; Os_delta – difference in outstanding between two consecutive periods; RLGD_os – realized LGD calculated with delta outstanding approach; Default_duration – duration of the default; Default_start_reason_X – reason of the default where X is the reason listed above and correspondingly means: 90 days past due, bankruptcy, failed probation period, fraud, restructurization, unlikeliness to pay, GDP – quarterly Gross Domestic Product, Employment – quarterly employment rate, HPI – quarterly House Price Index, Repayment – relative change in outstanding balance for reporting date$_t$ / reporting date$_{t-1}$; redefault – dummy variable where 1 means that it is not first default of this borrower.

Following Tong's et al. (2013) logic that we may benefit from previous loan balances, we included two variables particularly associated with the repayment dynamics. The first one is arbitral, that is delta outstanding amount (*Os_delta*). The second one is relative, that is relative change in outstanding balance between two neighboring periods – *repayment*. They also proposed utilization of information if the borrower has already defaulted in the past, which we included as *redefault* variable. Based on Ross and Shibut (2015) narration, we included the *Default_duration* variable, that is constant over the whole observation period per borrower. In the case of the multiple defaulters, we divided the default duration value for different defaults – the value will represent the exact duration of single default. In the case of draggings, we included the already known default duration, since complex models like XGBoost should be able to condition the weight of the variable with respect to the final status. Importantly for this research, which as a primary incentive deals with low data accessibility, all of mentioned variables: *Os_delta*, *repayment*, *default_duration* and *redefault*, are calculated based on the fundamental data about the credit: outstanding, interest rate, default date and reporting date.

Final status has been included in the model as a sample splitting factor already described in this section. This approach, i.e. using final status as a splitting factor instead of binary variable, will exclude the risk of bias originated from the fact that all exit no loss and cure cases have realized LGD = 0 at their last reporting date of the default. Moreover, variable correlations after the data split already reveal the contrast between correlation of RLGD_os with loss sample (0.889) RLGD and noloss sample (-0.159) RLGD. This may suggest unreliable results of delta outstanding approach for customers with final status "cured" or "not resolved". We analyze this issue further in the Section 4.1. Finally, it is worth to point out that some portfolios, due to the country of origin, have legal requirements to incorporate specific risk-drivers. Note that we excluded country-specific variables, since they would limit our research in terms of versatility of portfolios it can be applied to.

Our dataset contained also recovery and costs data, that are part of the cash-flow approach. For appropriate simulation of missing cash-flow data we have not included them in the model. It also includes *exit_balance* variable, which might have potentially introduced bias to the model. Given that each reporting date is treated independently, and the *exit_balance* remains constant throughout the entire period for a single default event, but target variable (realized LGD) fluctuates, inclusion of this variable would bring inconsistent information. To boot, we have not found any literature on including the information about the exit balance importance.

To summarize the variables selection, we have introduced loan-characteristic variables that are also common to use in literature for the LGD estimation and put our effort to obtain additional variables also supported by the literature using our fundamental data. On the other hand, we were limited with our dataset concerning the behavioral data, which are not included in the model. Finally, we excluded the cash-flow data, which according to our assumptions, are unavailable, and the data that may have potentially biased the model and have not been supported with the literature.

## 2.6 Training and evaluating the XGBoost model

To ensure the objectivity of our model's results, we first trained it on a dedicated training sample. Subsequently, we evaluated the model's performance using test samples, following a standard approach widely employed in the credit risk industry. This approach involves dividing the dataset into training, out-of-sample testing, and out-of-date testing subsets (Table 2). This method, as described by Tong et al. (2013), helps to validate the model's robustness over time and credit cycles, as well as prevent overfitting and improve its general reliability.

**Table 2.** Split for training and testing sample.

| Development period | | Out of date period |
|---|---|---|
| Training sample | Out of sample (test) | Out of date (test) |
| ca. 75% | ca. 20% | ca. 5% |

Note: Firstly, we excluded the last 6 months of observation period for the out of date sample, next, we performed traditional train-test split to obtain the 75% of data for training.

To appropriately compare the XGBoost model with delta outstanding approach we will extract only the observations within the test sample for the performance evolution of the delta outstanding approach.

To ensure the reliability and accuracy of our predictive model, we employed validation procedures (leaving the testing set untouched). Initially, we utilized a Grid Search Cross-Validation (GridSearchCV) approach on training sample to fine-tune the hyperparameters of our XGBoost Regressor model. GridSearchCV systematically explores a specified grid of hyperparameters and selects the combination that yields the best performance according to the chosen evaluation metric (loss function). In our case, the Mean Squared Error (MSE) was selected as the evaluation metric to optimize the model's predictive accuracy. However, due to the long computing time, we attempted to fine-tune the parameters using the RandomizedSearchCV algorithm. The procedure is similar and incorporates the same loss function, however it doesn't estimate the model for each set of hyperparameters grid, rather the random sets. Nevertheless, the crucial advantage of the RandomizedSearchCV was searching through the wider space of candidate hyperparameters in shorter time. In other words, adding the parameters to the grid will not decrease the efficiency of the algorithm, but widen the space of searching. The number of candidate parameters sets was controlled by *n_iter* parameters and in our case was set to 25, together with 5-fold cross-validation resulted in totaling 125 fits. After the estimation of the best performing XGBoost Regressor model using both algorithms, we could decide that we will utilize the parameters from RandomizedSearchCV, that have been on average performing better (Table 3). Note that the models for loss and noloss sample have been estimated separately and comparison in Table 3 is aimed to select the best fine-tuning approach, rather the model. For this reason, the loss and noloss cross-validation models are presented separately.

**Table 3.** Cross-validation results of all XGBoost models

| Level | Model | MAE | Sd. error | MSE |
|---|---|---:|---:|---:|
| **GridSearchCV** | | | | |
| Noloss | XGBoost | 0.004413 | 0.000410 | 0.000324 |
| Loss | XGBoost | 0.007125 | 0.000486 | 0.001015 |
| Total level | XGBoost | 0.014254 | 0.000551 | 0.002033 |
| **RandomizedSearchCV** | | | | |
| Noloss | XGBoost | 0.004528 | 0.000399 | 0.000316 |
| Loss | XGBoost | 0.006196 | 0.000413 | 0.000975 |
| Total level | XGBoost | 0.011505 | 0.000551 | 0.001835 |

Note: Validation algorithm included in GridSearchCV and RandomizedSearchCV with 5 folds. Both procedures perform cross-validation, splitting the train data further into the validation and train subsamples. Each iteration has one partition for testing (validating) and 4 partitions for training the model.

The hyperparameters grid we used in RandomizedSearchCV is presented in Table 4.

**Table 4.** Hyperparameters grid

| hyperparameter | Value |
|---|---:|
| learning_rate | [0.01, 0.05, 0.075, 0.1, 0.2] |
| max_depth | [7, 8, 9, 10, 11, 12, 13, 14, 15] |
| n_estimators | [700, 800, 900, 1000, 1100, 1150, 1200, 1250, 1300] |
| subsample | [0.6, 0.7, 0.75, 0.8, 0.85] |
| min_child_weight | [2, 3, 4, 5, 6] |
| colsample_bytree | [0.7, 0.8, 0.85, 0.9, 0.91, 0.92, 0.95] |

Note: learning_rate – parameter controlling the contribution of each tree to the ensemble; max_depth – max depth of each tree; n_estimators – number of trees in the model; subsample – fraction of samples used to train each tree; min_child_weight – minimum of samples required in each leaf node; colsample_bytree – controls fraction of features used to build each tree.

Hyperparameters levels after tuning are presented in Table 5.

**Table 5.** Hyperparameters after fine tuning.

| Hyperparameter/Sample | Noloss | Loss | All |
|---|---:|---:|---:|
| learning_rate | 0.05 | 0.1 | 0.05 |
| max_depth | 12 | 7 | 8 |
| n_estimators | 1200 | 1200 | 1300 |
| subsample | 0.80 | 0.85 | 0.80 |
| min_child_weight | 2 | 2 | 5 |
| colsample_bytree | 0.95 | 0.95 | 0.91 |

Note: Hyperparameters tuned with RandomizedSearchCV with following parameters: cross-validation folds: 5, number of interations: 25, n_jobs=-1 (use all available CPU cores for parallel processing), cost function: mean squared error, and parameters grid as presented in Table 4.

## 3. Results

In this section we present the results of estimated models and compare their performance with the results from the cash-flow approach. We employ performance metrics mentioned in the methodology section to analyze which model performed the best and to differentiate the performance depending on the sample used for the evaluation.

## 3.1 Delta outstanding approach vs. cash-flow approach

Since we fed the XGBoost model with the delta outstanding approach results, its accuracy is also an important factor, that might have either bias or boost the ML model's prediction power. Figure 5 represents the scatterplot of correlation between actual realized LGD value (based on cash-flows approach) with the delta outstanding approach results. Since the dataset on total level contains more than 50k observations, for better readability, the plot has been created based on random sample of 10k from the whole dataset.

**Figure 5.** Scatterplot of realized LGD estimated with delta outstanding approach vs cash-flow approach.

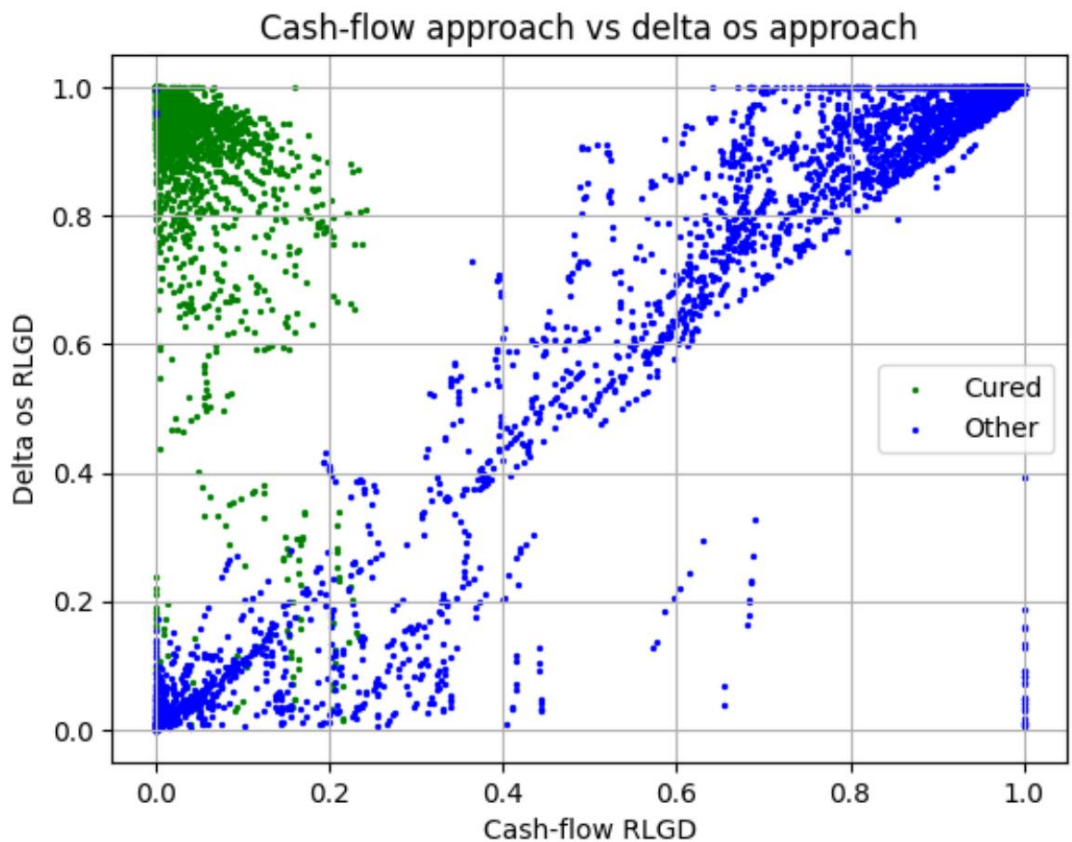

Note: We used random sampling with n=10 000 of observations out of total sample, where both cash-flow and delta os negative LGD's are capped to 0 (see, Equation 3) and LGD's exceeding 1 capped to 1. We did not control for the balance of final status in the random sample used for this plot. Based on own work.

Data points represented in Figure 5 reveal that the delta outstanding approach tendency of correctly approximated LGD is noticeable among prominent number of observations. For better comparability, relevant performance metrics presented in Table 6 were calculated for the subsets, representing the observations corresponding to the out of date and out of sample subsets of supervised models. However, MAE (0.227113) and MSE (0.187459) calculated on the total level are also introducing information for the research. Based on those statistics we can conclude, that the delta outstanding approach leaves a field for the improvement in estimating the realized LGD.

We also discovered that the bias observable on the scatterplot (overestimated LGD by the delta outstanding approach with relation to cash-flow approach) has been caused mainly by the cured borrowers. Transition from the defaulted to performing status (cure), depending on the bank's definition, generally requires completion of a probationary period associated with specified regular loan repayments (Wood and Powell, 2017). The delta outstanding approach based on the formula we applied is unable to capture the fact that the classification to the cure status is associated with final LGD equal to 0. From the perspective of delta outstanding approach, we observe the gradually decreasing outstanding amount (characteristic for curing cases) with no additional information of fulfillment for abovementioned condition, which results in resolved, cure case but with significant loss for the bank. Small, negative changes in outstanding, according to formula for Economic Loss noted in delta outstanding procedure (point 5), EL = os_ref – cumulative sum of delta os, will result in relatively high computer economic loss and in consequence, high RLGD, which is not reflected in the reality (RLGD calculated with cash-flow approach). For this reason, the methodology for delta outstanding procedure for cured cases is a potential further extension of this research.

**3.2 Benchmarking**

The training process, optimized with Randomized Search Cross-Validation, resulted with 3 best regressors for each set of data: noloss, loss and total sample. In Table 3, we presented the performance of loss and noloss sample separately to improve the interpretability and comparability of the model performance metrics in training and validation process. We used MSE and MAE metrics to compare the model's performance and they occurred to be consistent, even though only MSE was used as a cost function in the cross-validation. The best performance of the XGBoost regressor, measured with RandomizedSearchCV, is observed for noloss sample (MAE = 0.004528; sd. error = 0.000399; MSE = 0.000316). Subsequently, we observe less accurate regressor for loss sample (MAE = 0.006196; sd. error = 0.000413; MSE = 0.000975), and least accurate regressor for total sample (MAE = 0.011505; sd. error = 0.000551; MSE = 0.001835). Although the MAE and MSE metrics were used to compare the performance of the regressors, we also benefited from providing the standard deviation of errors. They reveal insights that Boost regressor for noloss sample was not only the most accurate, but also more stable in terms of predictions accuracy. The sd. errors values exabits similar increasing tendency as MSE and MAE, which mean that loss sample and total sample might have been more affected with the random fluctuations. However, the standard deviation of errors across various regressors shows minimal variation, providing little incentive to explore the underlying causes of prediction accuracy variability.

Table 6 contains the results of XGBoost performed on (1) the sample divided on loss and noloss customers, (2) the whole sample and results of delta outstanding approach. The metrics are calculated based on corresponding values of realized LGD calculated with cash-flow approach. The evaluation metrics contain mean absolute error, standard deviation of errors and mean squared error. The metrics are calculated for both test sample and out of date sample. Please note that the performance of models for No loss and Loss sample are presented jointly. This representation is motivated by the fact that the performance metrics in the below results are aimed to compare the models.

**Table 6.** Performance metrices for XGBoost models and delta outstanding approach.

| Level of application | Model | MAE | sd. error | MSE |
|---|---|---|---|---|
| **Out of sample** | | | | |
| No loss + loss | $XGB_{loss} + XGB_{noloss}$ | **0.005756** | **0.005756** | **0.000869** |
| Total level | $XGB_{total}$ | 0.010743 | 0.029428 | 0.000981 |
| Total level | Delta outstanding | 0.285511 | 0.355614 | 0.207978 |
| **Out of date** | | | | |
| No loss + loss | $XGB_{loss} + XGB_{noloss}$ | **0.021504** | **0.021504** | **0.002577** |
| Total level | $XGB_{total}$ | 0.090786 | 0.166008 | 0.035801 |
| Total level | Delta outstanding | 0.056838 | 0.213963 | 0.049011 |

Note: In 'Noloss + loss' level of application total number of predicted LGD's are the exactly the same as for Total level, but the XGBoost regressors are estimated separately on two subsamples, which is referred as '$XGB_{loss} + XGB_{noloss}$'. $XGB_{total}$ stands for XGBoost regressor estimated on the total training sample. The bolded values indicate the best results in each subperiods.

We observe that the performance is being improved when applying the XGBoost model onto the approximated LGD from delta outstanding approach. The performance of XGBoost trained on total level of data is particularly noticeable for the out of sample part. The main reason for weaker performance of XGBoost trained on total level of data and tested on out of date sample is that it consists mainly of unresolved cases (that limit the accessibility to whole customer's default history) (93.35%) and has minor participation of cured cases (4.98%), that were increasing the error of delta outstanding approach. Nevertheless, the MSE metric for XGBoost regressor on total level for out of date testing sample is lower than the metric for the delta outstanding approach on the out of date testing sample (0.035801 and 0.049011 correspondingly), which may indicate lower tendency of XGBoost regressor to return strongly outlying values.

Another improvement is observable after distinguishing the models for loss and noloss borrowers. Since both types of borrowers are expected to follow different patterns, we were able to calibrate the model more precisely. After this data transformation, two models trained separately on 'loss' and 'noloss' sample were jointly more accurate both for out of sample and out of date testing set. The higher consistency and confidence of the model predictions is also reflected in lower standard deviation of the errors for split model approach.

## 4. Sensitivity analysis

In this section we verify how robust are our results by varying the hyperparameters within a predefined range. The hyperparameters considered for the analysis include learning_rate, max_depth, n_estimators, subsample, min_child_weight, and colsample_bytree (presented in Table 4). To check the sensitivity of our results, we created a new set of hyperparameters for the wider range than in the grid used in model, increase the number of candidates sets and apply them into the RandomizedSearchCV. We also analyzed the sensitivity under two variable-related aspects: (1) if XGBoost regressor estimated on arbitral variables associated with delta outstanding approach (RLGD_os, outstanding, discount_factor) is worse than for the full set of variables (listed in Table 1) and (2) if XGBoost regressor estimated without information about result of the delta outstanding procedure (RLGD_os) is worse than for the full set of variables. For the better interpretability, we will use XGBoost on total sample.

### 4.1 Hyperparameters sensitivity

After a few iterations of assigning the grid, we found out that when selecting the random numbers from uniform distribution on the assumed interval, we benefit from limiting the interval to as close values as possible. Our new parameters grid is being created as follows: *n_estimators* – random 15 numbers from discrete uniform or uniform distribution on interval (900, 1400); *max_depth* – integers from 6 to 16; *learning_rate* - random 15 numbers from uniform distribution on interval (0.01, 0.2); *subsample* – random 15 numbers from uniform distribution on interval (0.6, 0.9); *colsample_bytree* – random 15 numbers from uniform distribution on interval (0.7, 0.95); *min_child_weight* – integers from 3 to 7. New and old parameters grid is summarized in Table 7.

**Table 7.** Hyperparameters after tuning for sensitivity analysis.

| Hyperparameter | N_estimators | Max_depth | Learning_rate | Subsample | Colsample_bytree | Min_child_weight |
|---|---|---|---|---|---|---|
| Value | 1337 | 9 | 0.081 | 0.868 | 0.861 | 7 |

Note: Hyperparameters tuned with RandomizedSearchCV with following parameters: cross-validation folds: 5, number of interations: 60, n_jobs=-1 (use all available CPU cores for parallel processing), cost function: mean squared error, and parameters grid as described in section *5.1 New hyperparameters and results.*

We performed RandomizedSearchCV gradually increasing the number of iterations by 5, up to the point, when algorithm will find the best set of hyperparameters. We succeeded to achieve better performance with 60 number of iterations, which together with 5-fold cross-validation resulted in totaling 300 fits. Comparison of performance is presented in Table 8.

**Table 8.** Performance metrices for first best XGBoost model and XGBoost model after sensitivity analysis for hyperparameters

| Level of application | Model | MAE | sd. error | MSE |
|---|---|---|---|---|
| **Cross-validation** | | | | |
| Total level | $XGB_{total}$ | 0.011505 | 0.000551 | 0.001835 |
| Total level | $XGB_{SA\_hyp}$ | 0.012387 | 0.000622 | 0.002007 |
| **Out of sample** | | | | |
| Total level | $XGB_{total}$ | 0.010743 | 0.029428 | 0.000981 |
| Total level | $XGB_{SA\_hyp}$ | 0.009882 | 0.028687 | 0.000921 |
| **Out of date** | | | | |
| Total level | $XGB_{total}$ | 0.090786 | 0.166008 | 0.035801 |
| Total level | $XGB_{SA\_hyp}$ | 0.092367 | 0.172867 | 0.038471 |

Note: $XGB_{total}$ is the model estimated based on the hyperparameters from **Table 5** and it is compared to $XGB_{SA\_hyp}$, model estimated on the hyperparameters from Table 7.

Our success in optimizing the hyperparameters can be attributed to two key factors:

1. **Iterative Optimization:** We employed an iterative approach, gradually increasing the number of candidates sets and iterations in the RandomizedSearchCV process. This systematic exploration allowed us to identify that the increasing number of iterations is beneficial for finding the best set of hyperparameters, but with a cost of computing power and time-consumption.

2. **Tailored Parameter Ranges:** Each hyperparameter was assigned specific ranges tailored based on the already estimated models.

In conclusion, our study demonstrates the effectiveness of systematic hyperparameter tuning in optimizing the performance of XGBoost models. By iteratively exploring the hyperparameter space and employing tailored parameter ranges, we successfully identified optimal configurations that significantly improved model performance. Our findings emphasize the importance of careful optimization process in machine learning model development and underscore the potential for further advancements through systematic parameter tuning.

## 4.2 Variables sensitivity

Although our primary focus is on exploring the challenges posed by limited access to cash-flow data, we extend our investigation to include an exploration of the implications of restricted access to other variables in the sensitivity analysis. This expanded scope allows us to dive deeper into the impact of data limitations on the performance of our model, offering valuable insights into its robustness and applicability in real-world scenarios.

**4.2.1 XGBoost for fundamental delta outstanding approach variables**

By restricting our model to input variables of delta outstanding approach (EAO, discount_rate) and its results (RLGD_os) we are able to assess sole impact of introduction of ML technique to our analysis. Holding the accessibility to the initial data unchanged, we can reflect the most restricted the conditions of the model developer.

For the possibly best interpretability, we maintained the parameters grid used in modeling part presented in Table 4. Comparison of cross-validation results for modeling part and sensitivity analysis part (regressor indexed with suffix SA_var1) are presented in Table 9.

**4.2.2 XGBoost for full variables set without the result of delta outstanding approach**

The foundational stage of our research lies in the calculation of the realized LGD using the delta outstanding approach, serving as the pivotal offset for subsequent analyses. One may say that our ML model was built to improve the approximation of the LGD approximation. To evaluate the importance of the delta outstanding step role in our model, we decided to compare the performance of two estimators. The first model utilizes the full set of variables, while the second model is constrained by the result of the delta outstanding approach.

For the possibly best interpretability, we maintained the parameters grid used in modeling part presented in Table 4. Comparison of the cross-validation results for modeling part and described sensitivity analysis part (regressor indexed with suffix SA_var2) are included in Table 9.

**Table 9.** Performance metrices for first best XGBoost model and XGBoost model after sensitivity analysis for variables

| Level of application | Model | MAE | sd. error | MSE |
|---|---|---|---|---|
| **Cross-validation** | | | | |
| Total level | XGB$_{total}$ | 0.011505 | 0.000551 | 0.001835 |
| Total level | XGB$_{SA\_var1}$ | 0.081478 | 0.002172 | 0.023343 |
| Total level | XGB$_{SA\_var2}$ | 0.017621 | 0.000685 | 0.002661 |
| **Out of sample** | | | | |
| Total level | Delta outstanding | 0.285511 | 0.355614 | 0.207978 |
| Total level | XGB$_{total}$ | 0.010743 | 0.029428 | 0.000981 |
| Total level | XGB$_{SA\_var1}$ | 0.080769 | 0.125434 | 0.022257 |
| Total level | XGB$_{SA\_var2}$ | 0.016674 | 0.036007 | 0.001574 |
| **Out of date** | | | | |
| Total level | Delta outstanding | 0.056838 | 0.213963 | 0.049011 |
| Total level | XGB$_{total}$ | 0.090786 | 0.166008 | 0.035801 |
| Total level | XGB$_{SA\_var1}$ | 0.322501 | 0.030897 | 0.199472 |
| Total level | XGB$_{SA\_var2}$ | 0.118895 | 0.068390 | 0.232925 |

Note: XGB$_{total}$ is the model estimated based on the hyperparameters from **Table 5** and it is compared to (1) XGB$_{SA\_var1}$ - model estimated on fundamental variables for delta outstanding approach (RLGD_os, EAO, discount_rate), (2) XGB$_{SA\_var2}$ - model estimated on full set of variables excluding RLGD_os.

### 4.2.3 Variables sensitivity conclusion

Considering the inherent characteristics of XGBoost models, applying strict constraints on variables may lead to suboptimal outcomes. This relation is observable in our findings, as the decrease in performance was stronger when applied on narrower set of variables. Notably, XGB$_{SA\_var1}$ consistently exhibits poorer performance compared to XGB$_{SA\_var2}$ across all sample sets (cross-validation, out of sample and out of date). Of particular significance is the observation that XGB$_{SA\_var1}$ outperformed the delta outstanding approach on the out-of-sample set. Despite the strongly restricted number of variables, there was a field for improvement for 'cured' cases that biased delta outstanding performance.

Conversely, XGB$_{SA\_var2}$ exhibited worse scores than XGB$_{total}$ across all sample sets. This finding suggests that excluding the delta outstanding procedure step, and consequently not incorporating its result into the model, could diminish the predictive power of the model.

### 5. Conclusions and discussion

Limited access to the data is a commonly encountered obstacle by the researchers, particularly in the domain of Credit Risk Modelling, where data availability significantly influences the precision of risk parameter estimation. This limitation primarily stems from the confidentiality and sensitivity of the data, as well as its magnitude or specificity. Even within the banking sector, cash-flow data faces various constraints, including portfolio-specific scenarios such as scope jumpers, time constraints forcing the analyst to perform collecting data process more general, or technical issues leading to missing data. To address the issue of missing cash-flow data, temporary mitigation measures can be implemented, particularly in processes like model monitoring, through the adoption of approaches such as the delta outstanding procedure.

In this research we used the dataset of retail mortgages from one of the European banks, that is characterized by completeness and access to the cash-flow data, on which we simulated the limited access to the cash-flow data. The dataset consists of 1891 borrowers, which were in default between the 2008 and 2019 and excludes the performing periods. Given that the data were reported monthly, it resulted with more than 56 thousand of single observations. Methodology implemented in this research uses the cash-flow approach of calculation realized LGD as the most robust way to determine its true value. We reflect the conditions of limited access to the cash-flow data and apply the delta outstanding approach to approximate the realized LGD. Furthermore, we assess the performance of the delta outstanding approach, applied with our best understanding of the description in the *Guide to Internal Models* published by the ECB. The supervised learning part adopts the methodology of splitting the data for training and testing samples. The training sample is cross-validated using the RandomizedSearchCV, which ensures the reliability of the results. The testing part is divided for the out of sample stage, where randomly selected testing data are used, and out of date stage, where we extract separate period in time, to provide a realistic assessment for the stability of the predictions across different time periods or on the future data. We primarily refer to the MAE and MSE performance metrics, commonly utilized in this type of models in the scientific literature. We also interpret the values of standard error, that explains the consistency of the predictions.

The assessment described in section 4.1 reveals that the method does not fully reflect the reality, when compared to the cash-flow approach (RQ1). Arguments against treatment the delta outstanding method as a perfect approximation of LGD are also presented in Table 6, where the performance of delta outstanding and supervised model is compared on the same subset of the data. Moreover, based on the relevant literature, we select variables commonly used to LGD models of mortgage loans and use the machine learning techniques to include them as the predictors for the realized LGD estimation. Our machine learning model has been trained using the best hyperparameters selected based on Randomized Search Cross-Validation. We discovered that inclusion of non-cash-flow related variables can lead to the improvement in performance of estimating the LGD (RQ2), as observed in Table 6 and Table 9. Finally, referring to methodologies described in the literature, we divided our dataset using the final status variable for loss and noloss subsamples to estimate two independent regressors. This modification resulted in better results of XGBoost model, according to employed performance metrics: MSE and MAE (RQ3), that are presented in Table 6.

Although the topic of limited access to exhaustive data for credit risk modeling is common in the literature, the amount of publications related to application of alternative ways of calculating realized LGD is not very representative. Elaboration on the delta outstanding approach contributes to the novel character of this research. Our study reveals that although the approach widely adopted in banks and endorsed by the ECB, exhibits inherent weaknesses. Its general methodology formulation leaves the space for the multiple interpretations and implementations among different institutions, and may yield imprecise or inconsistent results. We showed that generalized application of delta outstanding approach on all types of borrowers create a bias, misestimating the LGD of cured borrowers. Then, we presented that its efficiency can be enhanced through the application of machine learning techniques, resulting in significant improvements. Our approach represents a novel advancement in the field, demonstrating the efficacy of machine learning techniques in refining LGD approximation. By leveraging the XGBoost model's ability to incorporate a wider range of variables beyond traditional inputs like outstanding balance and interest rate, we achieved significant improvements in accuracy. This novel methodology allowed us to uncover more complex patterns within borrower behavior, enhancing our understanding of LGD dynamics.

We noted the XGBoost model estimated on total level sample improves the estimations in comparison to delta outstanding approach, when we evaluated the results on the testing set (out of sample). Nevertheless, the model was not sufficient for out of date testing sample, that contained the majority of unresolved cases. Our approach deviates from previous literature, which typically supports data segmentation based on customer current status (defaulted and performing). Nevertheless, our dataset contains only defaulted cases, that were subsampled based on their final status: (1) cured and exited without any loss for the bank and (2) not resolved and excited with loss. After such transformation of the data and estimation of two independent XGBoost regressors, joint performance of the models has been the best for both out of sample and out of date testing sets. This novel methodology underscores the significance of tailoring data segmentation strategies to suit specific dataset characteristics and contributes to advancing the understanding of effective modeling techniques in credit risk analysis.

Even though machine learning models are not yet commonplace in banking practices due to the challenges associated with interpretation, their theoretical potential suggests that certain banking procedures could benefit from their utilization. Further exploration within this domain may prompt questions regarding the efficiency of alternative data segmentation methods, or whether optimizing the delta outstanding model for cure cases poses a greater challenge in finding a more precise machine learning model. A primary area for further exploration involves improving the interpretation and application of the delta outstanding methodology for cured cases. Additionally, exploring the introduction of other models could offer valuable insights into modeling approaches tailored for conditions of limited data access.